\documentclass[a4paper,12pt]{article}
\usepackage{amsmath,amssymb,amsfonts,amsthm}
\newtheorem{theorem}{Theorem}[section]

\newtheorem{conjecture}[theorem]{Conjecture}
\usepackage{graphicx}

\newcommand{\gc}{g}
\newcommand{\nc}{\bar \nabla}
\newcommand{\vc}{\mathcal{V}}

\newcommand{\rc}{^{(4)}\mathcal R}

\newcommand{\norm}{\lambda}

\newcommand{\Y}{\omega}

\newcommand{\sliced}{S}

\newcommand{\gt}{h}

\newcommand{\vt}{\mathcal{N}}

\newcommand{\Nt}{n}

\newcommand{\dvq}{\, dV_q}

\newcommand{\x}{\sigma}

\newcommand{\mf}{\mathcal{M}}

\title{Angular momentum-mass inequality for axisymmetric black holes}
\author{Sergio Dain\\
  Facultad de Matem\'atica, Astronom\'{i}a y F\'{i}sica, \\
     Universidad Nacional de C\'ordoba, \\
     Ciudad Universitaria, (5000) C\'ordoba, \\
  Argentina}

\begin{document}
\maketitle
\begin{abstract}
In these notes we describe recent results concerning the inequality
$m\geq \sqrt{|J|}$ for axially symmetric black holes.  
\end{abstract}

\section{Introduction}
The following conjectures constitute the
essence of the current standard picture of the gravitational collapse:
i) Gravitational collapse results in a black hole (weak cosmic
censorship) ii) The spacetime settles down to a stationary final
state. If we further assume that at some finite time all the matter
fields have fallen into the black hole and hence the exterior region
is pure vacuum (for simplicity we discard electromagnetic fields in
the exterior), then the black hole uniqueness theorem implies that the
final state should be the Kerr black hole. The Kerr black hole is
uniquely characterized by its mass $m_0$ and angular momentum
$J_0$. These quantities   satisfy the following
remarkable inequality
\begin{equation}
 \label{eq:14i}
\sqrt{|J_0|}\leq m_0.
\end{equation}
From Newtonian considerations, we can interpret this inequality as
follows\cite{Wald71}: in a collapse the gravitational attraction
($\approx m_0^2/r^2$) at the horizon ($r \approx m_0 $) dominates over
the centrifugal repulsive forces ($\approx J_0^2/m_0r^3$).

If the initial conditions for a collapse violate \eqref{eq:14i} then
the extra angular momentum should be radiated away in gravitational
waves. However, in an axially symmetric spacetime the angular momentum is a
conserved quantity (the Komar integral of the Killing vector, see, for
example, \cite{Wald84}). In this case angular momentum cannot
be radiated: the angular momentum $J$ of the initial conditions must
be equal to the final one $J_0$. On the other hand, the mass of the
initial conditions $m$ satisfies $m\geq m_0$ because gravitational
radiation carries  positive energy. Then, from inequality
\eqref{eq:14i} we obtain
\begin{equation}
 \label{eq:14a}
\sqrt{|J|}\leq m.
\end{equation}
More precisely, i)-ii) imply that a complete, vacuum,
axisymmetric, asymptotically flat data should satisfy inequality
\eqref{eq:14a}, where $m$ and $J$ are the mass and angular momentum of
the data. Moreover, the equality in \eqref{eq:14a} should imply that
the data are a slice of extreme Kerr. This is a similar argument to
the one used by Penrose \cite{Penrose69} to obtain the inequality
between mass and the area of the horizon on the initial data. As in
the case of Penrose inequality, a counter example of \eqref{eq:14a}
will imply that either i) or ii) is not true. Conversely a proof of
\eqref{eq:14a} gives indirect evidence of the validity of i)-ii),
since it is very hard to understand why this highly nontrivial
inequality should hold unless i)-ii) can be thought of as providing
the underlying physical reason behind it (see the discussion in
\cite{Wald99}).

Inequality \eqref{eq:14a} is a property of the spacetime and not only
of the data, since both quantities $m$ and $J$ are independent of the
slicing.  It is in fact a property of axisymmetric, vacuum, black
holes space-times, because a non zero $J$ (in vacuum) implies a non
trivial topology on the data and this is expected to signal the
presence of a black hole. The physical interpretation of
\eqref{eq:14a} is the following: if we have a stationary vacuum black
hole (i.e. Kerr) and add to it axisymmetric gravitational waves, then
the spacetime will still have a (non-stationary) black hole, these
waves will only increase the mass and not the angular momentum of the
spacetime because they are axially symmetric.  Since inequality
\eqref{eq:14i} is satisfied for Kerr we get \eqref{eq:14a}.

In this note, we review some recent results (see \cite{Dain06c},
\cite{Dain05e}, \cite{Dain:2007pk}, \cite{Dain05d}, \cite{Dain05c},
\cite{Chrusciel:2007ak}) in which inequality \eqref{eq:14a} is proved
for one black hole and describe the open problems for the other cases.

\section{Variational principle for the mass}

Inequality \eqref{eq:14a}
suggests the following variational principle: 

\begin{quote}
\emph{The extreme Kerr
  initial data are the absolute minimum of the mass among all
  axisymmetric, vacuum, asymptotically flat and complete initial data
  with fixed angular momentum.}  
\end{quote}
However, it is important to note that for two related inequalities,
the positive mass theorem and the Penrose inequality, a variational
formulation was not successful. In the case of the positive mass
theorem only a local version was proved using a variational principle
\cite{Choquet-Bruhat76}.

The key difference in the present case is axial symmetry. As we will
see, in that case it possible to write the mass (in an appropriate
gauge) as a positive definite integral on a spacelike hypersurface.
The reason for this particular behavior of the mass is the
following. In the presence of a symmetry, vacuum Einstein equations
can be reduced a la Kaluza-Klein to a system on a 3-dimensional
manifold where it takes the form of 3-dimensional Einstein equations
coupled to a matter source.  Since in 3-dimension there is no
radiation (the Weyl tensor is zero), this source represents the true
gravitational degree of freedom that have descended from 4-dimensions
to appear as ``matter'' in 3-dimension.  Since all the energy is
produced by these effective matter sources, one would expect in that,
as in other field theories, the total energy of the system can be
expressed as a positive definite integral over them. This was in fact
proved by Brill \cite{Brill59} in some restricted cases and then
generalized in \cite{Gibbons06} \cite{Dain06c}\cite{Chrusciel:2007dd}.
Using this formula and with the extra assumption that the data are
maximal, the variational principle can be formulated in a very simple
form \cite{Dain05c}.

To write the mass formula for axially symmetric spacetimes we follow
\cite{Dain:2008xr}.  Consider a vacuum solution of Einstein's
equations, i.e., a four dimensional manifold $\vc$ with metric
$\gc_{ab}$ for which the Ricci tensor $ \rc_{ab}$ vanishes.  Suppose,
in addition, that there exists a spacetime Killing vector $\eta^a$. We
define the norm and the twist of $\eta^a$, respectively, by
\begin{equation}
  \label{eq:1}
\lambda^2=\eta^a\eta^b g_{ab}, \quad
\Y_a=\epsilon_{abcd}\eta^b \nc^c\eta^d,  
\end{equation}
where $\nc_a$ is the connection  and $\epsilon_{abcd}$ the volume
element with respect to  $\gc_{ab}$. Assuming that the manifold is
simply connected and using
$\rc_{ab}=0$ it is possible to prove that $\omega_a$ is the
gradient of a scalar field $\Y$
\begin{equation}
  \label{eq:2}
 \omega_a=\nc_a \Y. 
\end{equation}
In our case the Killing field will be spacelike, i.e. $\lambda \geq
0$. 

As we mention above, in the presence of a Killing field, there exists
a well known procedure to reduce the field equations \cite{Geroch71}.
Let $\vt$ denote the collection of all trajectories of $\eta^a$, and
assume that it is a differential 3-manifold.  We define the Lorentzian
metric $\gt_{ab}$ on $\vt$ by
\begin{equation}
  \label{eq:3}
\gc_{ab}=\gt_{ab}+ \frac{\eta_a\eta_b}{\norm^2}.
\end{equation}
Four dimensional Einstein vacuum equation are equivalent to Einstein
equations in three dimension on $\vt$ coupled to effective matter
fields determined by $\norm$ and $\Y$.  We make a $2+1$ decomposition
of these equations.  Let $\Nt^a$ be the unit normal vector orthogonal
to a spacelike, 2-dimensional slice $\sliced$.  The intrinsic metric
on $\sliced$ is denoted by $q_{AB}$ and the trace free part of the
second fundamental form of the slice is denoted by $k_{AB}$. On
$(\vt,\gt)$ we fix a gauge: the maximal-isothermal gauge (see
\cite{Dain:2008xr} for details) and the corresponding coordinates system
$(t,\rho, z)$. It is convenient
to define the function $\x$ by
\begin{equation}
  \label{eq:9}
  \norm=\rho e^{\x/2}. 
\end{equation}

In this gauge the mass can be written in the following form 
\begin{equation}
  \label{eq:15b}
  m= \frac{1}{16}\int_{S}\left(
    2k^{AB}k_{AB}+3\frac{\norm'^2}{\norm^2}  +  \frac{ \Y'^2}{\norm^4}
    +|D\x |^2 + \frac{ |D\Y|^2}{\norm^4}
  \right )\rho \dvq. 
\end{equation}
where $\dvq=e^{2u}d\rho dz$ denote the volume element with respect to
$q_{ab}$, $D$ is the covariant derivative with respect to $q_{AB}$
with $|D\x |^2= D^A\x D_A\x $, and the prime denotes directional
derivative with respect to $n^a$, that is
\begin{equation}
  \label{eq:14}
\norm' =n^a \nc_a \norm.
\end{equation}
This is essentially a derivative  with respect to $t$. 
Note that all the terms in the integrand of (\ref{eq:15b}) are
positive definite. The first three terms contain the dynamical part of
the data, they vanish for stationary solutions, in particular for the
Kerr solution. The last two terms, contain the stationary part of the
fields. It is important to note that the integral of these terms does
not depends on the metric $q_{AB}$. In effect, the integral of these
terms can be written as
\begin{equation}
  \label{eq:5c}
 \mf(\x,\Y)= \frac{1}{16}\int_{-\infty}^{\infty}dz \int_0^\infty d\rho
  \left(|\partial  \x |^2  +\rho^{-4} e^{-2\x} |\partial \Y |^2
  \right) \rho,
\end{equation}
where $\partial$ denotes partial derivatives with respect to $(\rho, z)$.
The integral (\ref{eq:5c}) depends only on $\x$ and $\Y$. Since we have
\begin{equation}
  \label{eq:5}
  m\geq \mf,
\end{equation}
to find the minimum of $m$ is equivalent as to find the minimums of
$\mf$. 

\begin{figure}
\label{fig:1}
\begin{center}
\includegraphics[width=4cm]{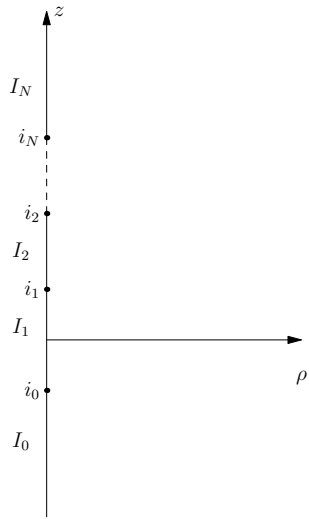}
\caption{$N$ asymptotic ends}
\end{center}
\end{figure}

In order to write the variational principle, it only remains to
discuss the boundary conditions. Physically, we want to prescribe
boundary conditions such that the total angular momentum is fixed. The
information of the angular momentum is determined by the value of the
twist potential $\Y$ at the axis $\rho=0$ (see \cite{Dain05c}). To include more
than one black hole, we prescribe the following topology.  Let $i_k$
be a finite collection of points located at the axis $\rho=0$.  Define
the intervals $I_k$, $0\leq k\leq N-1 $, to be the open sets in the
axis between $i_k$ and $i_{k-1}$, we also define $I_0$ and $I_N$ as
$z< i_0$ and $z >i_N$ respectively. See figure \ref{fig:1}. Each point
$i_k$ will correspond to an asymptotic end of the data, and hence we
will say that the data contain $N$ black holes. 

To fix the total angular momentum $J$ (where $J$ is an arbitrary
constant) of the data is equivalent as to
prescribe the following boundary condition for $\Y$ (see \cite{Dain06c})
\begin{equation}
  \label{eq:7}
  \Y|_{I_0}=4J, \quad  \Y|_{I_N}=-4J. 
\end{equation}

We want to study the minimums of the functional $\mf$ with these
boundary conditions.

We are now in position to write the precise form of the variational
principle. 
\begin{conjecture}
\label{con:1}
  Let $\x, \Y$ be arbitrary functions such that $\Y$ satisfies
the boundary   condition \eqref{eq:7}. Then we have
  \begin{equation}
    \mf (\x,\Y)\geq \sqrt{|J|}. 
  \end{equation}
Moreover, the equality implies that $\x,\Y$ are given by the extreme
Kerr solution.
\end{conjecture}

This conjecture was proved for the case $N=1$ in \cite{Dain06c}.  This
result was extended in \cite{Chrusciel:2007ak} to include more generic
data. 

The conjecture is open for the case $N\geq 2$. For this case, 
 the variational problem is fixed if we impose
the boundary condition
\begin{equation}
  \label{eq:8}
   \Y|_{I_i}=4J_i,
\end{equation}
with $0<i<N$, for arbitrary constants $J_i$. Note however, that
conjecture \ref{con:1} is independent of the values $J_i$.

Remarkably, in \cite{Chrusciel:2007ak} it is proved that the
variational problem has a solution (i.e. a minimum) for arbitrary $N$,
but the value of $\mf$ for this solution is not known. In order to
prove the conjecture for $N\geq 2$, one need to compute a lower bound
for this quantity.  This problem is related  with the uniqueness of
the Kerr black hole with degenerate and disconnected horizons.


\end{document}